# Measuring Spreadsheet Formula Understandability

Felienne Hermans, Martin Pinzger & Arie van Deursen
Mekelweg 4, Delft
{f.f.j.hermans, m.pinzger, arie.vandeursen}@tudelft.nl

## ABSTRACT

*Spreadsheets are widely used in industry, because they are flexible and easy to use. Sometimes they are even used for business-critical applications. It is however difficult for spreadsheet users to correctly assess the quality of spreadsheets, especially with respect to their understandability. Understandability of spreadsheets is important, since spreadsheets often have a long lifespan, during which they are used by several users. In this paper, we establish a set of spreadsheet understandability metrics. We start by studying related work and interviewing 40 spreadsheet professionals to obtain a set of characteristics that might contribute to understandability problems in spreadsheets. Based on those characteristics we subsequently determine a number of understandability metrics. To evaluate the usefulness of our metrics, we conducted a series of experiments in which professional spreadsheet users performed a number of small maintenance tasks on a set of spreadsheets from the EUSES spreadsheet corpus. We subsequently calculate the correlation between the metrics and the performance of subjects on these tasks. The results clearly indicate that the number of ranges, the nesting depth and the presence of conditional operations in formulas significantly increase the difficulty of understanding a spreadsheet.*

## 1 INTRODUCTION

Spreadsheets are used extensively in business, for all sorts of tasks and purposes. While other assets of companies---like software products and processes---are strongly guarded, spreadsheets are usually not structurally checked. This lack of control contrasts their impact, which can be very large, as previous studies have shown. For instance, Hall [Hall, 1996]interviewed 106 spreadsheet developers and found that only 7% of the spreadsheets were of low importance and that as much as 39% were of high importance. In a more recent study we found similar results [Hermans, 2011].

Another observation we did in previous research on spreadsheet use, is that spreadsheets often stay in use for several years, in which their complexity increases gradually, due to repeated addition of data and the manipulation of formulas. Spreadsheet users lack the knowledge to decide whether the quality and structure of the spreadsheets remain within reasonable borders during its lifespan. The lifespan of a spreadsheet can furthermore involve several users. In an industrial case we found spreadsheets that had been use since 1997, and spreadsheets that were used by as many as 60 employees[Hermans, 2011]. These findings underline the importance of spreadsheet understandability and maintainability.
This paper determines a set of understandability metrics for spreadsheet formulas that support end-users in deciding whether their spreadsheet is *healthy*. With the metrics we aim to predict to what extent future users will be able to understand the spreadsheet.
We establish the spreadsheet metrics by studying related work and by interviewing spreadsheet users to identify a set of possible spreadsheet characteristics contributing to comprehension problems. From those two sources we distilled a number of metrics, which are divided in two categories: formula complexity: relating to the length and structure of formulas, and formula placement: concerning the location of formulas relative to their references. Those metrics are subsequently evaluated by an empirical study, in which we gather the metrics for 15 randomly selected formulas from the EUSES spreadsheet corpus[Fisher, 2005]. This corpus consists of 5,000 real-life spreadsheets from domains such as finance, biology and education. We determine whether the metrics influence understandability by letting 40 spreadsheet professionals analyze the 15 formulas and perform three tasks on them: 1) assessing their own understanding on a four point scale; 2) explaining the formula; and 3) finding the cells that the formula refers to.

From this empirical analysis we learn that metrics in the two metric categories show significant correlations with understandability. With respect to complexity, the number of ranges, the nesting depth and the presence of conditional operations in formulas are of influence. In the placement category, formulas are experienced as more difficult to understand when they refer to cells outside of their own row and to cells that are located to the right and below the formula.

## 2 APPROACH & OBJECTIVES

### 2.1 Objectives

In previous work in the area of spreadsheets and spreadsheet comprehension we have seen that it is difficult for end-users to assess the complexity of spreadsheets they worked with. Experiments show that even users who were considered experts on a certain spreadsheet had trouble correctly assessing its complexity[Hermans, 2011].

The inability of end-users to correctly determine the complexity of spreadsheets leads to several problems. Firstly, when a complex spreadsheet is transferred to a colleague or has to be adapted this will be more time-consuming when a spreadsheet is complex. Since the lifetime of spreadsheets can be as long as 15 years[Hermans, 2011], the use of complex spreadsheets costs companies considerable amounts of time. Secondly, it has been shown before that complex spreadsheets make error finding more difficult[Teo, 2001] and that complex cells have a higher potential for faults than others[Panko, 1998]. In this paper we aim at determining which spreadsheets formula characteristics make spreadsheets hard to understand. We focus on the technical aspects of spreadsheets, since they are easy to analyze automatically, and it is possible to control them. There are,of course, other aspects that might influence spreadsheet comprehension, such as the time that was spent on the creation of the spreadsheet or the spreadsheet knowledge of the creator. These aspects are generally hard to measure and quantify, and even harder to change, and are therefore left out of the scope of this paper.

Thus, the aim of our research is to *understand how spreadsheet formula characteristics affect understandability*.

### 2.2Approach

To establish a set of metrics, we use two sources. Firstly, we interview a number of spreadsheet professionals, asking them what their most prevalent problems with spreadsheets are. Secondly, we investigate related work in the area of spreadsheets. We subsequently refine the research question into measurable spreadsheet formula metrics.

After the establishment of the metrics, we evaluate them. Therefore we perform an empirical study, in which 40 spreadsheet professionals evaluated 15 spreadsheet formulas. For each of the formulas, participants are asked to express their understanding on a four point scale, explain the formula in words and identify all the cells the formula refers to. We subsequently correlate the performance of subjects on these tasks with the metrics to determine the relation between the metrics and understandability.

## 3 ESTABLISHING THE METRICS

As stated above, to obtain the metrics, we use two sources. Firstly, we conducted interviews with 40 spreadsheet professionals, using reputational case selection. Reputational selection involves using community experts suggesting the best informants, based upon what the researcher wants to study [Schensul, 1999].

### 3.1 Participants

The participants are all employees of the Robeco. Robeco is a Dutch asset management company with approximately 1600 employees worldwide, of which 1000 work in their headquarters in Rotterdam, where we performed our experiments. The following gives an overview of the participants in the interviews and in the subsequent experiment.

- All participants were frequent Excel users, using spreadsheets at least one time a week.
- The age of the participants varied from 24 to 51, with an average of 35.
- Their experience with Excel spreadsheets varied from 2 years up to as many as 20 years, with an average of 9.4.
- We asked participants to classify their Excel level themselves, using the classification from Hole *et al.* [Hole, 2009].
  - o 50% of all participants classified themselves as Basic Excel Users, meaning they use Excel for storing lists and repetitive calculations and have some experience with functions, pivot tables and charts.
  - o The other 50% of subjects named themselves Power User, meaning having a wide understanding of Excel's functionality, being able to create complex spreadsheets for own use and helping colleagues to develop and debug spreadsheets.
- Only 19% of participants had ever worked with other spreadsheet systems than Excel; all of those named Lotus 1-2-3 as the system they worked with.

### 3.2 Interviews

Our interviews were focused on answering the following question: 'What are the characteristics of a spreadsheet that make it hard to understand?'. We gathered the answers of all interviews, and found that many answers were similar. The following list shows the most important factors in understanding a spreadsheet that were named in the interviews, many of which also appear in related work.

- Formulas with many references, especially those scattered over the spreadsheet
- References between worksheets
- Formulas containing nested calculation [Bregar, 2004]
- Conditional formula constructions [Hodnigg, 2008]

This categorization of answers gives an overview of the factors contributing to difficulties in understanding spreadsheets, and thus helps in defining the detailed research questions. Related work on spreadsheet design and spreadsheet metrics gave rise to additional ideas.

The first aspect gathered from related work is placement of formulas, which is often addressed in papers on spreadsheet design guidelines, such as[Read, 1999] and [Rajalingham, 2000]. References right or below a formula are often missed by spreadsheet users, which is why the guidelines advice against it. Furthermore the length of calculation chain matters [Bregar, 2004]. When users have to step through a long list of formulas to find references, the change of missing one is probably higher, making it more difficult to understand and maintain. Finally we learned from related work that the distance between formula and referenced cells might influence understandability [Panko, 1998]. When references are distant, the change that users might miss them is higher, reducing understandability. This adds the following spreadsheet characteristics to the list.

- Directionality of the formulas
- Length of calculation chain
- Distance between formula and referenced cells

When compiling this list of metrics, we noticed that they can be divided into two categories. Some metrics (like formulas with many references) relate to the complexity of formulas, while others concern the placement, like the distance between a formula and its references.

## 4 METRICS

### 4.1 Formula Complexity

Many participants in the questionnaire named complex formulas as troublesome, so we will measure various complexity characteristics. Firstly, we count the number of references a formula has (M1.1). We also take the grouping of those references into account, as inspecting one range (e.g., SUM(A1:A10)) differs from inspecting a formula that is constructed of individual cells A1+A2+...+A9+A10, even though they results in the same answer (M1.2).

Related work [Bregar, 2004], [Read, 1999] describes that formula construction involving logic and conditions are considered more difficult for users than calculation formulas (M1.3).
Furthermore these papers name nested formulas as hard, something that has been confirmed by our interviewees. We will measure nestedness of a formula as the height of the parse tree of the formula (M1.4).

Related work finally suggests that a long calculation chain increases the change of errors[Bregar, 2004]. A calculation chain is defined as a number of cells that are linked by formula references. With the length of the calculation chain we mean the number of cells one has to traverse to reach the referenced cell that lays farthest away (M1.5).

- M1.1 The number of direct references
- M1.2 The number of ranges in which the references are grouped. This is a value between 1, when all references are contained in one range, and the number of references, when all cells are referred individually.
- M1.3 The presence of conditional operations in a formula
- M1.4 Nestedness of formulas: measured as the height of the parse tree
- M1.5 Length of calculation chain

## 4.2 Formula Placement

From related work [Read, 1999] we learn that references to cells located to the left or above of the formula are considered easier to comprehend than those referring right or down. We therefore introduce a metric determining the percentage of references that is *reverse*, meaning located to the right of, or below of the formula (M2.1). Cells lying in a different worksheet are considered reverse if they lay in a worksheet right of the worksheet the formula lays in. Interviewees mentioned that they consider formulas difficult when the references are scattered over the worksheet. They reckon that it is easier when all references are either in the same row (M2.2) or in the same column (M2.3), since it is then easy to oversee all of them at once.

Finally, we measure formula distance (M2.4). We suspect it gets more difficult to quickly understand a formula when its precedents lay geographically further away, since that involves scrolling, or even inspecting different worksheets. For this metric we define cells that are *distant* as cells that lie further away than 10 columns or 25 rows from the formula. We chose these numbers, since this is approximately the number of rows and columns that will be visible when a user inspects a formula. References in different worksheets are also considered to be distant. The above leads to the following formula placement metrics.

- M2.1 Percentage of reverse references
- M2.2 Percentage of references in the same row
- M2.3 Percentage of references in the same column
- M2.4 Percentage of distant references

## 5 SPREADSHEET FORMULAS FOR THE STUDY

### 5.1 Setup

As a source for the experimental phase we used 15 formulas stemming from spreadsheets of the EUSES Spreadsheet Corpus [Fisher, 2005]. This corpus consists of around 5000 spreadsheets, divided over 11 categories---ranging from educational to financial---and has been used by several researchers to evaluate algorithms on spreadsheets, among which [Hermans, 2010] and[Abraham, 2006]. We selected spreadsheets from the financial folder only, since this is the domain our subjects are most familiar with. By selecting spreadsheets from a known domain, we diminish the influence of domain knowledge on the results. The financial folder of the EUSES corpus consists of 781 spreadsheets. From this folder we randomly chose five spreadsheets and from each of those five spreadsheets, we randomly selected three formulas, for which we analyzed formula and understandability metrics. We chose five spreadsheets and used 3 of their formulas, rather than choosing 15 formulas from 15 different spreadsheets, because it would be too time consuming for subjects to investigate 15 different spreadsheets within the tests. The formulas we used can be found in Table 1. All 5 spreadsheets can be downloaded from our research page (http://swerl.tudelft.nl/bin/view/FelienneHermans/Publications).

| # | Location | Formula |
|---|----------|---------|
| 1 | D29 | (((C4+C5)+C24)+(E15-E14))-(C15-C14) |
| 2 | D4 | IF(('Input Information'!E19=0),"Must input data",((('Input Information'!C32+'Input Information'!C30)-'Input Information'!C26)/'Input Information'!E19)) |
| 3 | C17 | IF((C5=0),"Must input data",((C10+C13)/C5)) |
| 4 | N38 | SUM(N31:N37) |
| 5 | D21 | SUM(D18:D19) |
| 6 | N42 | SUM(N40:N41) |
| 7 | G16 | G11+G15 |
| 8 | C46 | C23+C33+C44+C45 |
| 9 | D40 | D36/D38 |
| 10 | F18 | SUM(D10:D17) |
| 11 | F33 | F29-F31 |
| 12 | D94 | SUM(C54:C94) |
| 13 | E11 | IF(B11>0,LN(D11),) |
| 14 | E47 | (((EXP((0-G36)*C36))*C34*C42-C35*(EXP((0-G34)*C36))*C45)-((EXP((0-G36)*C36))*C34*F42-C38*(EXP((0-G34)*C36))*F45))*C37/G37 |
| 15 | C41 | (LN(C34/C35)+(G34-G36+(G35/2))*C36)/(((G35)^(0.5))*(C36^0.5)) |

| # | M1.1 | M1.2 | M1.3 | M1.4 | M1.5 | M2.1 | M2.2 | M2.3 | M2.4 |
|---|------|------|------|------|------|------|------|------|------|
| 1 | 7 | 7 | 0 | 4 | 1 | 28 | 0 | 71 | 28 |
| 2 | 5 | 5 | 1 | 5 | 3 | 100 | 0 | 0 | 100 |
| 3 | 4 | 4 | 0 | 4 | 4 | 0 | 0 | 100 | 0 |
| 4 | 7 | 1 | 0 | 1 | 1 | 0 | 0 | 100 | 0 |
| 5 | 2 | 1 | 0 | 1 | 3 | 0 | 0 | 100 | 0 |
| 6 | 2 | 1 | 0 | 1 | 1 | 0 | 0 | 100 | 0 |
| 7 | 2 | 2 | 0 | 1 | 2 | 0 | 0 | 100 | 0 |
| 8 | 4 | 4 | 0 | 1 | 8 | 0 | 0 | 100 | 25 |
| 9 | 2 | 2 | 0 | 1 | 6 | 0 | 0 | 100 | 0 |
| 10 | 8 | 1 | 0 | 1 | 1 | 0 | 0 | 100 | 0 |
| 11 | 2 | 2 | 0 | 1 | 3 | 0 | 0 | 100 | 0 |
| 12 | 41 | 1 | 0 | 1 | 3 | 0 | 2 | 0 | 48 |
| 13 | 2 | 2 | 1 | 2 | 2 | 0 | 100 | 0 | 0 |
| 14 | 18 | 18 | 1 | 6 | 5 | 38 | 0 | 0 | 0 |
| 15 | 8 | 8 | 1 | 4 | 2 | 50 | 0 | 50 | 0 |

# 6 EVALUATION

## 6.1 Setup

To evaluate the chosen metrics, we performed an empirical study in which we invited the 40participants of the initial interviews to inspect a set of spreadsheet formulas.

In the experiments we measure two different forms of understandability. Firstly we measure perceived understandability, by asking participants how well they understand the formulas. For this we use the following four point scale of levels of understandability:

- It is easy: I understand it well and could change it
- It is somewhat easy: I can work with it, but I would rather not change it
- It is difficult: I have an idea what happens, but I could not modify it
- It is very difficult: I have no clue how this formula works

We furthermore asked participants to explain why they selected the level of understanding, by providing them with a free text field to elaborate on their choice.

Besides this perceived understanding, we measure the understandability in an experimental way, by means of two tasks participants had to perform on the formula.

First, we ask them to explain the calculation in the formula in words. With this we assess whether participants understand the calculation in the formula cell. We score this explanation on a four point scale as follows:

- We score an answer as correct earning four points if both the right operations are named (like: sum, if, minus) and also the meaning of the formula in terms of the spreadsheet's context is correct (like: this is a balance, this is a gross profit etcetera)
- When only the right formula was named, the score is three.
- When only the right meaning is named, score is two points.
- In case both are missing, or the answer was something completely different, a participant scores one point.

For the second experimental measure subjects are asked to list all cells the formula references. We explained to users in the beginning of the study, that they were supposed to list all cells the formula depended on, also references on other sheets, or indirectly referenced cells. Since understanding spreadsheets often involves understanding formula structure, this task is an indication for the ability of users to understand the formula. This assignment is also scored in four categories, defined as follows:

- If a subject names all references, including those further along the calculation chain, a participant earns four points.
- When some of the references along the chain are entered, but not all of them, the score is three.
- If only direct references were named, the score is two
- Finally, if references were missing or superfluous, the score is one.

We measure these two forms of understandability---perceived and experimental--- to gain a more thorough evaluation of understandability, for it is possible that some of the metrics influence perceived understandability rather than the actual ability to perform maintenance tasks. If the lay-out of a spreadsheet is nice this could lead the user to think he understands the formulas very well, while the structure is not actually improving understandability. With this, we have the following three measures for understandability:

- U1Perceived understandability
- U2Experimental understandability: explanation
- U3 Experimental understandability: reference finding

## 6.2 Participants

In the experiment the subjects were the same 40 that we previously interviewed. Before the start of the experiment, we did not reveal the metrics to the subjects, in order not to bias them in any way. Robeco does not have any spreadsheet best practices in use, nor did subjects follow any Excel courses where they might have learned them. So, other than their own personal preferences, subjects were unaware of spreadsheet best practices.

## 6.3 Results

### 6.3.1 Understandability

Figure 1 shows the median of the results of all 40 participants, on both perceived understandability (U1), as well as the performance on the two formula tasks, describing the formula (U2), and analyzing the references (U3). We see that the ability to explain a formula and its perceived complexity are more or less in line: It seems users are able to assess their understanding of the meaning of the formula quite well.

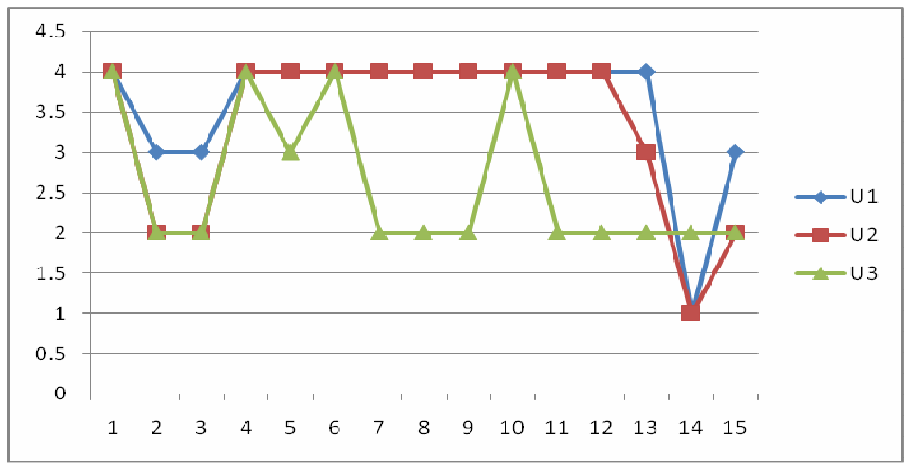

**Figure 1 The medians of all 40participants' scores on the three understandability measures for the 15 randomly selected formulas**

In Figure 1 we furthermore see that the relation between perceived understandability (U1) and the score for finding all the formula references (U3), is more capricious. Apparently users have a tendency to overestimate their ability to understand and manipulate a formula. For instance in the case of formulas 5, 7, 8, 9, 11, and 12, the median of the scores on U1 is 4, meaning 'This formula is easy: I understand it well and I could change it'. However the median of the scores on U3, reference finding, is only 2, indicating most participants could only identify those references located on the same worksheet as the formula. The fact that subjects do not have a clear overview poses a serious threat to their ability to manipulate the formula, especially since they themselves feel that they do have good insight into the formula. This overconfidence of spreadsheet users has been addressed earlier [Panko, 1998].

This fact underlines the added value of automated analysis of spreadsheets. There are clearly formula characteristics that are overlooked by spreadsheet users, even by those who consider themselves spreadsheet experts.

### 6.3.2 Metrics

With the metrics of the 15 formulas and the understandability results of the 40 professionals, we revisit our research question: how do spreadsheet formula characteristics affect understandability? To do so, we determine the influence of the individual metrics of perceived understandability and experimental understandability, i.e. we calculate the Spearman correlation of the values of M1.1 to M2.4 with the scores for U1 to U3.

#### 6.3.2.1 Complexity metrics

Table 3 SPEARMAN CORRELATION BETWEEN THE UNDERSTANDABILITY MEASURES AND FORMULA COMPLEXITY METRICS M1.1 TO M1.5 significant correlations are marked with ** for the 0.01 level or * for the 0.05 level

|    | M1.1   | M1.2      | M1.3      | M1.4      | M1.5      |
|----|--------|-----------|-----------|-----------|-----------|
| U1 | -0.227 | -0.631*   | -0.646**  | -0.845**  | -0.233    |
| U2 | -0.309 | -0.693**  | -0.600*   | -0.884**  | -0.274    |
| U3 | 0.181  | -0.359    | -0.302    | -0.198    | -0.785**  |

To determine the influence of the five complexity metrics on understandability, we perform a correlation analysis on the data. In this correlation analysis, we calculated the Spearman rank between the formula metrics and the medians of the understandability measures for all 40 participants of the experiments. Table 3 lists the Spearman rank for the five formula complexity metrics and the three understandability measures. In this table, as in subsequent tables, significant correlations at the 0.01 level are marked with **, and those significant at 0.05 level are indicated with *. As can be seen in Table 3, there are three metrics that affect both the perceived and the experimental understandability considerably. Firstly there is the number of ranges (M1.2). The influence of a low number of ranges on understandability is supported by quotes of participants, who expressed their preference for formulas that "reference only one group of cells, which makes it easier to see what is being done". Secondly, there is the strong influence of conditional operations (M1.3). Related work [Hodnigg, 2008] assessed that conditionals are notoriously difficult for end users. Our experiment confirms this, both by a strong statistical correlation, and by users, who stated in our experiment that conditional operations are "a different type of formulas that are not simple" and that "it took me some time to understand ifs and vlookups, this is not something you grow up with, like sum and multiply. And I still need to look twice when I see one". Nesting depth (M1.4) also matters, about which we gathered from a participant: "A nice formula is simple, and contains only one operation. I don't like it when multiple operations are combined into one formula". The total number of references (M1.1) does not correlate with any of the understandability metrics. From this, combined with the above stated, we can conclude that it is important to group references in as few ranges as possible, but not necessarily to limit the number of references when striving for understandable formulas. Another interesting fact is the influence of the length of the calculation chain. This seems to only affect the scores on U3, reference finding. We conclude that users find it difficult to work with long calculation chains, which, however, does not influence their perceived understanding of the formula or their ability to explain it.

Summarizing the above, we conclude that formula complexity indeed influences understandability of spreadsheet formulas. In particular the number of ranges, the nesting depth, the presence of logical operations, and a long calculation chain has a large influence on user comprehension.

### 6.3.2.2 Placement metrics

Table 4 SPEARMAN CORRELATION BETWEEN THE UNDERSTANDABILITY MEASURES AND FORMULA PLACEMENT METRICSM2.1TOM2.3significant correlations are marked with ** for the 0.01 level or * for the 0.05 level

|  | M2.1 | M2.2 | M2.3 | M2.4 |
|---|---|---|---|---|
| U1 | -0.560* | -0.220 | 0.753** | 0.344 |
| U2 | -0.653** | -0.047 | 0.720** | 0.344 |
| U3 | -0.090 | -0.236 | 0.325 | 0.075 |

For the second metric category, we again performed a Spearman correlation between the four formula location metrics and the three understandability metrics. As shown in Table 4, there are two metrics in the formula placement category that significantly correlate with the understandability metrics; The number of reverse references (M2.1) influences understandability measures U1 and U2 weakly, and a stronger correlation is found with the percentage of references in the same column (M2.3).

The numbers show that users feel formulas are easier and they are able to explain and find references better when all references are in the same column. Participants confirmed this in their free text assessment of the formulas: "It is better when the linked cells are neatly structured, which makes working with the formula easier" and "I hate it when I have to look for all the corresponding cells, because they are in different places".

From these results we can conclude that formula placement affects understandability, though less strongly than formula complexity. Both reverse references and especially references in the same column influence the capability of subjects to understand a formula.

### 6.4 Conclusion

The aim of this study was to understand how spreadsheet formula characteristics affect understandability. Based on the results we have presented above, we can conclude that

- Formula complexity does influence understandability. Especially the **number of ranges**, the **nesting depth**, the presence of **logical operations** and a **long calculation chain**.
- Formula placement affects understandability to a lesser extent. **Reverse references** and especially **references in the same column** influence the capability of subjects to understand a formula.

## 7. DISCUSSION

### 7.1 High Level Metrics

In the current approach we measure metrics at the level of formulas. It would however be very insightful to determine and measure metrics at higher levels, like at the level of worksheets or spreadsheets. Interesting metrics at those levels could include the number of different formula cells within a worksheet, or the number of worksheets in a spreadsheet file. In future work we will investigate the influence of these high level metrics in detail. We do however believe that formulas are the basic ingredient of a spreadsheet, and hence formula metrics are very important metrics for spreadsheet understandability.

### 7.2 Application perspective

The current set of metrics could be used by spreadsheet intensive organizations to determine whether the spreadsheets they have in use are understandable and maintainable. The metrics indicate whether certain formulas are understandable, so they could be used to direct users to formulas that could be revised. The applicability of the metrics will be addressed in a bigger industrial case study in the future.

## 8 CONCLUDING REMARKS

The aim of this paper is to develop a set of metrics that indicate the understandability of a give spreadsheet. To that end we have studied related work and interviewed spreadsheet professionals resulting in a set of metrics. These metrics were evaluated by collecting them for a set of fifteen spreadsheet formulas randomly chosen from the EUSES spreadsheet corpus, and measuring the understandability of professional spreadsheet users, both by asking them directly, as by testing it with assignments on the formulas. We conclude that formula complexity metrics, in particular the number of ranges, the nesting depth and a long calculation chain. To a lesser extent the placement of formulas relative to their references also influences understandability, especially the percentage of references in the same column as the formula aids in understanding.

The key contributions of this paper are as follows:

- A survey analyzing key factors that influence spreadsheet comprehension.
- A list of spreadsheet metrics resulting from this survey
- An evaluation of the metrics within the context of large Dutch financial company
- A thorough analysis of the result of this evaluation.

The current research gives rise to several directions for future work. Firstly, it would be interesting to investigate metrics on higher levels of granularity, like data blocks, worksheets or spreadsheet files, also taking the connectivity of worksheets into account. Furthermore the influence of layout factors, like coloring, borders or fonts on understandability is an interesting avenue for further research. Last but not least it would be interesting to investigate how the metrics evolve over the years that a spreadsheet has been in use.

## 9 ACKNOWLEDGEMENTS

The authors would like to thank all the participants of the tests at Robeco Rotterdam for their participation and valuable feedback.